\documentclass[aps,prb,twocolumn,showpacs]{revtex4}
\usepackage{amssymb}
\usepackage{amsmath}
\usepackage{epsfig}

\begin{document}

\title{Visibility of current and shot noise in electrical Mach-Zehnder and Hanbury Brown Twiss interferometers}

\author{V.S.-W. Chung$^{1,2}$, P. Samuelsson$^{3}$, and
M. B\"{u}ttiker$^{1}$} 
\affiliation{$^{1}$D\'{e}partement de Physique
Th\'{e}orique, Universit\'{e} de Gen\`{e}ve, Gen\`{e}ve 4, CH-1211
Switzerland\\ $^{2}$Department of Electronics, National Chiao-Tung
University, Hsinchu 30010, Taiwan\\ $^3$Division of Solid State
Theory, Lund University, S\"olvegatan 14 A, S-223 62 Lund, Sweden}
\date{\today }

\begin{abstract}
We investigate the visibility of the current and shot-noise
correlations of electrical analogs of the optical Mach-Zehnder
interferometer and the Hanbury Brown Twiss interferometer. The
electrical analogs are discussed in conductors subject to high
magnetic fields where electron motion is along edge states. The
transport quantities are modulated with the help of an Aharonov-Bohm
flux.  We discuss the conductance (current) visibility and shot noise
visibility as a function of temperature and applied voltage. Dephasing
is introduced with the help of fictitious voltage probes.  Comparison
of these two interferometers is of interest since the Mach-Zehnder
interferometer is an amplitude (single-particle) interferometer
whereas the Hanbury Brown Twiss interferometer is an intensity
(two-particle) interferometer.  A direct comparison is only possible
for the shot noise of the two interferometers.  We find that the
visibility of shot noise correlations of the Hanbury Brown Twiss
interferometer as function of temperature, voltage or dephasing, is
qualitatively similar to the visibility of the first harmonic of the
shot noise correlation of the Mach-Zehnder interferometer. In
contrast, the second harmonic of the shot noise visibility of the
Mach-Zehnder interferometer decreases much more rapidly with
increasing temperature, voltage or dephasing rate.
\end{abstract}

\maketitle

\section{Introduction}

With the advent of mesoscopic physics, it has become possible to
experimentally investigate quantum phase coherent properties of
electrons in solid state conductors in a controlled way. In
particular, in ballistic mesoscopic samples at low temperatures,
electrons can propagate up to several microns without loosing phase
information. This opens up the possibility to investigate electrical
analogs of various optical phenomena and experiments. An investigation
of such analogs is of fundamental interest. On the one hand, it allows
one to establish similarities between the properties of photons and
conduction electrons, a consequence of the wave nature of the quantum
particles. On the other hand, it also allows one to investigate the
differences between the two types of particles arising from the
different quantum statistical properties of fermions and bosons. For
many-particle properties, such as light intensity correlations or
correspondingly electrical current correlations, noise, the quantum
statistical properties are important. \cite{Buttiker92,Blanter00} Both
the wave-nature of the particles as well as their quantum statistics
are displayed in a clearcut fashion in interferometer structures. In
this work we are concerned with the electrical analogs of two well
known optical interferometers, the single-particle Mach-Zehnder (MZ)
interferometer and the two-particle Hanbury Brown Twiss (HBT)
interferometer.

The MZ-interferometer is a subject of most textbooks in
optics. \cite{BornWolf} In the framework of quantum optics, considering
individual photons rather than classical beams of light, the
interference arises due to the superposition of the amplitudes for two
different possible paths of a single photon. This leads to an
interference term in the light intensity. The MZ-interferometer is
thus a prime example of a single particle
interferometer.\cite{Mandel99} Various electronic interferometers with
ballistic transport of the electrons have been investigated
experimentally over the last decades, as e.g. Aharonov-Bohm (AB)
rings\cite{ABrings} and double-slit interferometers.\cite{Buks99}
Detailed investigations of dephasing in ballistic interferometers was
carried out in Refs. [\onlinecite{Hansen01,Kobayashi02}]. Only very
recently was the first electronic MZ-interferometer realized by Ji et
al.\cite{Ji03} in a mesoscopic conductor in the quantum Hall regime. A
high visibility of the conductance oscillations was observed, however
the visibility was not perfect. This led the authors to investigate in
detail various sources for dephasing. As a part of this investigation,
also shot noise was measured. Still, some aspects of the experiment
are not yet fully understood. Theoretically, Seelig and one of the
authors\cite{Seelig01} investigated the effect of dephasing due to Nyquist 
noise on the conductance in a
MZ-interferometer. The effect of dephasing on the closely related
four-terminal resistance in ballistic interferometers \cite{Seelig03}
was investigated as well. Dephasing in ballistic strongly interacting systems 
is discussed by Le Hur. \cite{Lehur02,Lehur05} 
Following the experimental work of Ji et
al.,\cite{Ji03} Marquardt and Bruder investigated the effect of
dephasing on the shot-noise in MZ-interferometers, considering
dephasing models based on both
classical\cite{Marquardt04a,Marquardt04b} as well as quantum
fluctuating fields. \cite{Marquardt04c} Very recently, F\"orster, Pilgram 
and one of the authors \cite{Forster05} extended the dephasing model of
Refs. [\onlinecite{Seelig01,Marquardt04a}] to the full statistical
distribution of the transmitted charge.

The HBT-interferometer \cite{HBT1,HBT2,HBT3} was originally invented
for stellar astronomy, to measure the angular diameter of stars. It is
an intensity, or two-particle,\cite{Mandel99} interferometer. The
interference arises from the superposition of the amplitudes for two
different two-particle processes. Importantly, there is no single
particle interference in the HBT-interferometer. Consequently, in
contrast to the MZ-interferometer there is no interference in the
light intensity, the interference instead appears in the
intensity-intensity correlations. Moreover, the intensity-intensity
correlation also display the effect of quantum statistics. Photons
originating from thermal sources tend to bunch, giving rise to
positive intensity cross correlations. For the electronic analog of
the HBT-interferometer, it was the corresponding anti-bunching of
electrons that originally attracted interest. It was predicted
\cite{Buttiker92} that the electrical current cross correlations in
mesoscopic conductors would be manifestly negative, i.e. display
anti-bunching, as a consequence of the fermionic statistics of the
electrons. Negative current cross correlations were subsequently
observed in two independent experiments.\cite{Oliver99,Henny99}
Recently, anti-bunching for field emitted electrons in vacuum was also
demonstrated.\cite{Kiesel02} The two-particle interference in the
HBT-experiment has received much less attention. We emphasize that
while the bunching of the photons was necessary for obtaining a
finite, positive cross correlation signal, it was the two-particle
effect that was of main importance to HBT since the angular diameter
of the star was determined from the two-particle interference pattern.
In electrical conductors, two-particle effects in AB-interferometers
were investigated theoretically in
Refs. [\onlinecite{Butt91,Butt92b,Loss00}]. Only very recently two of
the authors and Sukhorukov\cite{Sam04} proposed a direct electronic
analog of the optical HBT-interferometer which permits to demonstrate
two-particle interference in an unambiguous way.

In this work we investigate and compare in detail the current and and
zero-frequency noise in electronic MZ and HBT interferometers. We
consider interferometers implemented in mesoscopic conductors in the
integer Quantum Hall regime, where the transport takes place along
single edge states and Quantum Point Contacts (QPC's) serve as
controllable beam splitters. The effect of finite temperature, applied
bias and asymmetry, i.e. unequal length of the interferometer arms, is
investigated. The strength of the interference contribution is
quantified via the visibility of the phase oscillations. The
dependence of the visibility on the beam splitter transparencies as
well as on the temperature, voltage and asymmetry is discussed in
detail. Of interest is the comparison of visibility of the shot-noise
correlation of the MZ-interferometer and the HBT-intensity
interferometer.  Shot noise correlations in the MZ-interferometer
exhibit two contributions, one with the fundamental period of $h/e$
and a second harmonic with period $h/2e$. The shot noise correlations
in the HBT-interferometer, even though they are due to two particle
processes, are periodic with period $h/e$. Thus the Aharonov-Bohm
period can not be used to identify the two particle processes which
give rise to the HBT effect. It is therefore interesting to ask
whether the HBT two-particle processes have any other signature, for
instance in the temperature or voltage dependence of the visibility of
the shot-noise correlation.  We find that this is not the case. To the
contrary, we find that the the shot noise correlations in the HBT
intensity interferometer behave qualitatively similar to the $h/e$
shot noise correlation in the MZ-interferometer.  In contrast the
$h/2e$ contribution in the shot noise of the MZ-interferometer
decreases more rapidly with increasing temperature, voltage or
dephasing rate than the $h/e$ oscillation in the MZ- or
HBT-interferometer.

We investigate dephasing of the electrons propagating along the
edge states by connecting one of the interferometer arms to a
fictitious, dephasing voltage probe. In all cases, the current and
noise of the MZ-interferometer as well as the noise in the
HBT-interferometer, the effect of the voltage probe is equivalent to
the effect of a slowly fluctuating phase.

\section{Model and Theory}

\subsection{Optical analogs in the Quantum Hall regime}

In the paper we consider implementations of the MZ and HBT
interferometers in mesoscopic conductors in strong magnetic fields, in
the integer Quantum Hall regime.\cite{Klitzing80} The typical system
is a two-dimensional electron gas in a semiconductor heterostructure,
with the lateral confinement of the electron gas controllable via
electrostatic gating.  The transport between
reservoirs\cite{Buttiker88} connected to the conductor takes place
along edge states.\cite{Halperin82} The edge states, quantum analogs
of classical skipping orbits, are chiral, the transport along an edge
state is unidirectional. Scattering between edge states is suppressed
everywhere in the conductor except at electrostatically controllable
constrictions, QPC's.\cite{vanWees88,Wharam88} For a magnetic field
that does not break the spin degeneracy of the edge states, each edge
state supplies two conduction modes, one per spin.

These properties make conductors in the integer quantum Hall regime
ideal for realizing analogs of optical experiments. First, the edge
states correspond to single mode waveguides for the light. The
unidirectional motion along the edge states allows for ``beams'' of
electrons to be realized. Second, the QPC's work as electronic beam
splitters with controllable transparency. Moreover, due to chirality
the beamsplitters are reflectionless, a property essential for the MZ
and HBT interferometers but difficult to achieve for beam splitters in
conductors in weak (or zero) magnetic fields.\cite{Liu98,Oliver99}
These properties of conductors in the quantum Hall regime have been
demonstrated experimentally in a number of works, see
e.g. [\onlinecite{Henny99,Ober00,Ji03}].

Theoretically, several works have been concerned with the conductance
and noise properties of beam splitters and interferometers in Quantum
Hall systems, for a recent reviews see
e.g. Refs. [\onlinecite{Blanter00,But03}]. Recently, it was proposed
to use these appealing properties of edge states in the context of
orbital \cite{Sam03} quasi-particle entanglement in static
\cite{Been03,Sam04,Been04} and dynamic \cite{Sam04b,Been05} systems as
well as for quantum state transfer.\cite{Stace04}

It is interesting to note that the edge state description also hold
for conductors at even higher magnetic fields, in the fractional
Quantum Hall regime. As examples, the fractional charge has been
determined in shot-noise experiments\cite{Samin97,Picc97} and the
quantum statistical properties of the fractionally charged
quasi-particles have been investigated theoretically in
beam-splitter\cite{Safi01} and HBT\cite{Vish03} geometries. Various
interferometer structures have also been
considered.\cite{Kivel89,Chamon97,Geller97} Very recently, a
MZ-interferometer in the fractional Quantum Hall regime was
proposed.\cite{Jonck05} In this work we however consider only the
integer Quantum Hall effect, where the quasi-particles are
noninteracting and the electrical analogs to optical experiments can
be directly realized.

\subsection{Scattering approach to current and noise}

This discussion leads us to consider single mode, multi-terminal
conductors with noninteracting electrons. The principle aim of this
work is a comparison of the MZ and HBT-interferometers. In reality in
both interferometers interactions (screening) play a role both for the
voltage and temperature dependence. A non-interacting scattering
approach is not gauge invariant but requires a treatment of screening.
\cite{tcmb} However, these effects are expected to be simliar in the
two interferometers and will not affect the main conclusions of this
work. Therefore, below we treat non-interacting qausi-particle
interferometers.  The conductors are connected to several electronic
reservoirs, biased at a voltage $eV$ or grounded. The
current\cite{But86} and the noise\cite{But90,Buttiker92} are
calculated within the scattering approach for multi-terminal
conductors.  We first introduce the creation and annihilation
operators for ingoing, $\hat{a}_{\alpha }^{\dagger }(E)$ and
$\hat{a}_{\alpha }(E)$, and outgoing, $\hat{b}_{\alpha }^{\dagger
}(E)$ and $\hat{b}_{\alpha }(E)$, particles, at energy $E$ in terminal
$\alpha $. For simplicity we suppress spin notation. Considering a
conductor with $N$\ terminals, the in- and out-going annihilation\
operators are related via the $N\times N$ scattering matrix, as
\begin{equation}
\hat{b}_{\alpha }(E)=\sum_{\beta =1}^{N}s_{\alpha \beta }(E)\hat{a}_{\beta
}(E)
\end{equation}
where $s_{\alpha \beta }(E)$\ is the amplitude to scatter from terminal $
\beta $ to terminal $\alpha $. The current operator in the lead $\alpha $
has the form\cite{But86}
\begin{eqnarray}
\hat{I}_{\alpha }(t)&=&\frac{e}{h}\sum_{\beta \gamma }\int dEdE^{\prime
}\mbox{exp}(i[E-E']t/\hbar) \nonumber \\
&\times& 
A_{\beta \gamma }^{\alpha
}(E,E^{\prime}) \hat{a}_{\beta }^{\dagger}(E)\hat{a}_{\gamma
}(E^{\prime}),
\end{eqnarray}
with the notation
\begin{equation}
A_{\beta \gamma }^{\alpha }(E,E^{\prime })=\delta _{\alpha \beta }\delta _{\alpha
\gamma }-s_{\alpha \beta }^{\ast
}(E)s_{\alpha \gamma }(E^{\prime }).
\label{Adef}
\end{equation}
The average current is given by\cite{But86} 
\begin{equation}
\left\langle I_{\alpha }\right\rangle =\int dEj_{\alpha }(E),
\label{avcurr}
\end{equation}
where the spectral current density is
\begin{equation}
j_{\alpha }(E)=\frac{1}{e}\sum_{\beta }G_{\alpha \beta }(E)f_{\beta }(E).
\label{currspecdens}
\end{equation}
Here $f_{\beta }(E)=1/(1+\exp \left[ (E-eV_{\beta})/k_{B}T\right])$ is
the Fermi Dirac distribution of terminal $\beta $, with $V_{\beta}$
the corresponding applied voltage. The spectral conductance $G_{\alpha
\beta }(E)$ is given by
\begin{equation}
G_{\alpha \beta }(E)=\frac{e^{2}}{h}A_{\beta \beta }^{\alpha }(E,E).
\label{speccond}
\end{equation}
The zero frequency correlator between current fluctuations in terminals $%
\alpha $ and $\beta $\ is defined as 
\begin{equation}
S_{\alpha \beta }=\int dt \langle \Delta \hat{I}_{\alpha }\left(
0\right) \Delta \hat{I}_{\beta }\left( t\right) +\Delta \hat{I}_{\beta
}\left( t\right) \Delta \hat{I}_{\alpha }\left( 0\right) \rangle,
\end{equation}
where $\Delta \hat{I}_{\alpha }\left( t\right) =\hat{I}_{\alpha }\left(
t\right) -\langle {\hat{I}_{\alpha }\left( t\right) }\rangle .$
The current correlator is given by\cite{But90,Buttiker92} 
\begin{equation}
S_{\alpha \beta }=\int dES_{\alpha \beta }(E)
\label{noise}
\end{equation}
where
\begin{eqnarray}
S_{\alpha \beta }(E) &=&\frac{2e^{2}}{h}\sum_{\gamma \delta }A_{\gamma
\delta }^{\alpha }(E,E)A_{\delta \gamma }^{\beta }(E,E)  \notag \\
&\times&f_{\gamma }(E)\left[ 1-f_{\delta }(E)\right]
\label{noisedens}
\end{eqnarray}
is the spectral current correlator.

\subsection{Dephasing voltage probe model}

There are several physical mechanisms that might lead to dephasing of
the electrons propagating along the edge states (see e.g. the
discussion in Ref. [\onlinecite{Ji03}]). In this work we are not
interested in any particular mechanism for dephasing but consider
instead a phenomenological model, a dephasing voltage probe. The idea
of using a voltage probe to induce dephasing was introduced in
Refs. [\onlinecite{Butt86,Butt88b}]. A voltage probe connected to a
mesoscopic sample was considered, leading to a suppression of coherent
transport due to inelastic scattering. The probe model, originally
considered for the average current, was extended to treat the effect
of inelastic scattering on shot noise by B\"uttiker and
Beenakker\cite{Been92} by considering a conservation of current
fluctuations at the probe as well. Later De Jong and
Beenakker\cite{Jong96} extended the voltage probe concept and
introduced a (fictitious) voltage probe which breaks phase but does
not dissipate energy. Scattering in the voltage probe is
(quasi-)elastic.  This is achieved with the help of a distribution
function in the voltage probe which conserves not only total current
like a real voltage probe, but conserves current in every small energy
interval. Such a probe provides a model of pure dephasing.  The
different probe models have been used as qualitative models in a
number of works, see Refs. [\onlinecite{Langen97,Blanter00}] for a
review. For an application to quantum Hall systems, see
Ref. [\onlinecite{Texier00}].

In this work we consider the dephasing voltage probe model, which
conserves the current at each energy. The model is based on the
assumption that the current is conserved on a time scale $\tau_C$,
much shorter than the time of the measurement but much longer than the
time between injection of individual electrons, here of the order of
$\hbar/eV$. One could however consider a more general voltage probe
model that takes into account a more complicated dynamics of the
probe. A detailed discussion of such a general model in the light of
recent work\cite{Marquardt04a,Marquardt04b,Clerk04,Been05b} is however
deferred to a later work. Here we only note that below we find that
the voltage probe in both the MZ and HBT-interferometers only gives
rise to a suppression of the phase dependent terms in conductance and
noise, just as one would naively expect to be the effect of pure
dephasing.

The condition of zero current into the fictitious probe $\gamma$ at
each energy is fulfilled by considering a
time dependent distribution function of the probe
\begin{equation}
f_{\gamma}(E,t)=\bar{f_{\gamma }}\left(E\right) +\delta f_{\gamma}(E,t),
\end{equation}
where $\delta f_{\gamma}(E,t)$ fluctuates to conserve current on the
timescale $\tau_C$. As a consequence, the spectral current density at
each energy in lead $\alpha$ fluctuates in time as
\begin{equation}
j_{\alpha }(E,t)=j_{\alpha }(E)+\Delta j_{\alpha }(E,t),
\end{equation}
where the fluctuations $\Delta j_{\alpha }(E,t)=\delta j_{\alpha
}(E,t)+(1/e)G_{\alpha\gamma}(E)\delta f_{\gamma }(E,t)$ consist of two
parts, the intrinsic fluctuations $\delta j_{\alpha }(E,t)$ and the
additional fluctuations due to $\delta f_{\gamma }(E,t)$. The
requirement of zero average current into the probe, $j_{\gamma
}(E)=0$, leads to the averaged distribution function at the probe
reservoir $\gamma$
\begin{equation}
\bar{f_{\gamma}}\left( E\right) =-\sum_{\alpha \neq \gamma }\frac{G_{\gamma
\alpha }(E)}{G_{\gamma \gamma }(E)}f_{\alpha }(E).
\label{fav}
\end{equation}
The average spectral current density $j_{\alpha}^{dp}(E)$ is then
found from Eq. (\ref{currspecdens}).
 
The fluctuating part of the distribution function, $\delta f_{\gamma
}(E,t),$ is obtained from the requirement of zero current fluctuations
into the probe, $\Delta j_{\gamma }(E,t)=\delta j_{\gamma
}(E,t)+\left( 1/e\right) G_{\gamma \gamma }(E)\delta f_{\gamma
}(E,t)=0.$ The total current density fluctuation is then given by
\begin{equation}
\Delta j_{\alpha }(E,t)=\delta j_{\alpha }(E,t)-\frac{G_{\alpha \gamma }(E)}{
G_{\gamma \gamma }(E)}\delta j_{\gamma }(E,t).
\end{equation}
As a result, in the presence of dephasing the total spectral current
correlation $S_{\alpha \beta }^{dp}(E)$ is
\begin{eqnarray}
S_{\alpha \beta }^{dp}(E) &=&S_{\alpha \beta }(E)-\frac{G_{\alpha
\gamma }(E) }{G_{\gamma \gamma }(E)}S_{\beta \gamma
}(E)-\frac{G_{\beta \gamma }(E)}{G_{\gamma \gamma }(E)}S_{\alpha
\gamma }(E) \notag \\ &+&\frac{G_{\alpha \gamma }(E)G_{\beta \gamma
}(E)}{G_{\gamma \gamma }^{2}(E)}S_{\gamma \gamma }(E),
\label{dephnoisedens}
\end{eqnarray}
where $S_{\alpha \beta }(E)$ is the correlation function between the
intrinsic current fluctuations, $\delta j_{\alpha }$ and $\delta
j_{\beta }$, of contact $\alpha $\ and $\beta $, given by
Eq. (\ref{noisedens}), and $G_{\alpha \beta}(E)$ is the conductance,
given by Eq. (\ref{speccond}).
\begin{figure}[t]
\includegraphics[width=0.45 \textwidth,angle=0]{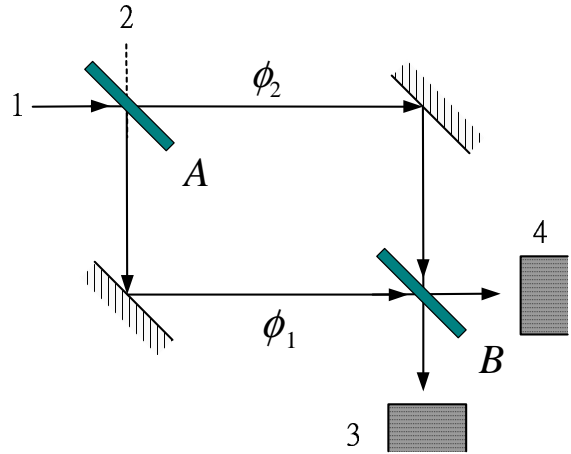}
\caption{An optical Mach-Zehnder interferometer. A beam of light
incident from $1$ is split in two partial beams at the semitransparent
beam splitter $A$. The two partial beams acquire geometrical phases
$\phi_1$ and $\phi_2$ respectively and are rejoined at the second beam
splitter $B$. The light intensity is measured in detectors $3$ and
$4$}
\label{MZopt}
\end{figure}

\section{Mach-Zehnder interferometers}

A schematic of the MZ-interferometer is shown in Fig. \ref{MZopt}. An
incident beam of light from source $1$ is divided in two parts at the
semitransparent beam splitter $A$. The two partial beams are reflected
at mirrors and later joined at the second beam splitter B.  Beams of
light going out from $B$ are detected in $3$ and $4$. The amplitude of
the light in an outgoing beam is the sum of the amplitudes for the two
partial beams, $A=A_1\mbox{exp}(i\phi_1)+A_2\mbox{exp}(i\phi_2)$. This
gives an intensity
$|A|^2=|A_1|^2+|A_2|^2+2\mbox{Re}\left\{A_1A_2^*\exp(i[\phi_1-\phi_2])\right\}$. The
interference term
$2\mbox{Re}\left\{A_1A_2^*\exp(i[\phi_1-\phi_2])\right\}$ thus
contains the difference between the geometrical phases,
$\phi_1-\phi_2$. Importantly, the four terminal geometry together with
the reflectionless beam splitters lead to that the incident beam
traverses the interferometer only once. This is a defining property of
the MZ-interferometer.
\begin{figure}[t]
\includegraphics[width=0.4 \textwidth,angle=0]{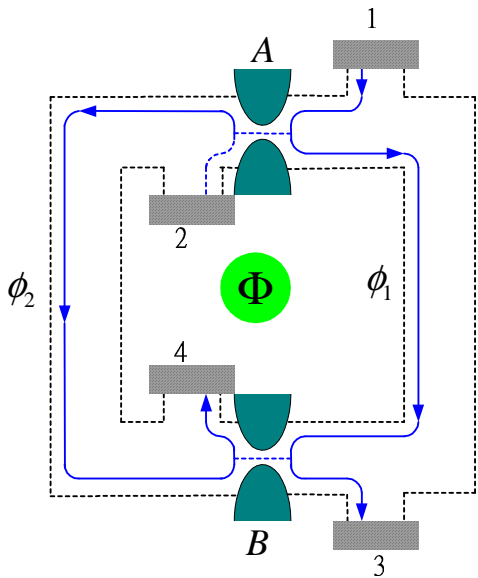}
\caption{The electronic analog of the MZ-interferometer, implemented by 
Ji et al. \cite{Ji03} 
in a conductor in the Quantum Hall regime. The electronic reservoir
$1$ is biased at $eV$ and reservoirs $2$ to $4$ are kept at
ground. The edge states (solid lines) have a direction of transport
indicated by arrows. The QPC's $A$ and $B$ play the role of the
beam splitters in Fig. \ref{MZopt}. Geometrical phases $\phi_1$ and
$\phi_2 $ and the AB-flux $\Phi$ are shown.}
\label{MZel}
\end{figure}

We then turn to the electric analog of the MZ-interferometer, shown in
Fig. \ref{MZel}. As pointed out above, several results for the current
and noise are available in the
literature.\cite{Seelig01,Seelig03,Marquardt04a,Marquardt04b,Marquardt04c,Forster05}
Here we analyze the most general situation possible, with finite
voltage, temperature and interferometer arm asymmetry as well as
different beam splitters $A$ and $B$ with arbitrary transparency. When
we consider limiting cases for e.g. small temperature, bias or
asymmetry, known results are recovered. This detailed analysis of the
MZ-interferometer is of importance when comparing to the
HBT-interferometer below.

We first discuss a fully coherent interferometer, the effect of
dephasing is investigated below.  An electric potential $eV$ is
applied at terminal 1, all other terminals are kept at zero
potential. The injected electrons propagate along single edge states. 
Scattering between the edge states can take place only at the two
QPC's, acting as beam splitters with controllable transparency. The
beam splitters $j=A,B$ are characterized by the scattering matrices

\begin{equation}
\left( 
\begin{array}{cc}
i\sqrt{R_{j}} & \sqrt{\mathcal{T}_{j}} \\ 
\sqrt{\mathcal{T}_{j}} & i\sqrt{R_{j}}
\end{array}\right),
\label{smat}
\end{equation}
where $\mathcal{T}_{j}$ and $R_{j}=1-\mathcal{T}_{j}$ are the
transmission and reflection probabilities, respectively. We note that
any additional phases of the beam splitters just give rise to a
constant phase shift of the oscillations in the interference terms and
are therefore not considered.

Propagating along the edge states, the electrons pick up geometrical phases
$\phi_{1}$\ and $\phi_{2}$ as well as phases $\psi_{1}$ and $\psi
_{2}$ due to the AB-flux $\Phi$ through the center of the
interferometer. For example,
the amplitude for scattering from terminal 1 to 4 is given by
\begin{equation}
s_{41}=i\sqrt{\mathcal{T}_{B}R_{A}}e^{i(\phi _{1}+\psi _{1})}+i\sqrt{
\mathcal{T}_{A}R_{B}}e^{i(\phi_{2}-\psi _{2})}
\label{s41}
\end{equation}
For the geometrical phases, to be specific we consider the case when
the potential landscape $eU(x,y)$ of the conductor in Fig. \ref{MZel}
is varying smoothly on the scale of the magnetic length
$l_B=(\hbar/e|B|)^{1/2}$, with $B\hat z$ the applied magnetic field
perpendicular to the plane in Fig. \ref{MZel} (the effect of
selfconsistency of the potential \cite{chshg} is
neglected). This allows for a semiclassical treatment. \cite{Fertig}
In a high magnetic field the edge states at Fermi energy $E_F$ follow
equipotential lines determined by $eU(x,y)=E_F-\hbar\omega_c (n+1/2)$
where $\omega_c=eB/m$ is the cyclotron frequency and $m$ the effective
electron mass. We are concerned here with the case where there is only
one edge state and thus $n = 0$. Suppose the $x$-axis is a line
intersecting quantum point contacts $A$ and $B$ in
Fig. \ref{MZel}. Excluding self-intersections we can express the edge
state in terms of functions $y_{1}(x)$ and $y_{2}(x)$ for the left and
right path of the interferometer. Working in the symmetric gauge, the
geometric phases can be written \cite{Fertig} $\phi_i=-l^{-2}_B
\int_{x_A}^{x_B} dx y_i (x)$, where $x_A$ and $x_B$ are the locations
of the QPC's. Importantly, $\phi_1-\phi_2$ corresponds to the total
area $A$ enclosed by these two paths divided by the magnetic length
squared, or $\phi_1-\phi_2=2\pi BA/\Phi_0$ where $BA$ is the total
flux through the enclosed area and $\Phi_{0}=h/e$ the elementary flux
quantum. Note that the Aharonov-Bohm flux $\Phi$ adds an additional
phase $\psi _{1}$ and $\psi_{2}$, with $\psi
_{1}+\psi_{2}=2\pi\Phi/\Phi_{0}$, to each of the paths.

For the discussion of the temperature and voltage dependence of the
current and the noise, we also need to know the energy dependence of
the phases. First, instead of parameterizing the edge state through $x$
we introduce the parameter $s$ which measures directly the path
length, i.e. $x(s)$, $y(s)$. In addition at $s$ we introduce local
coordinates $s_{\parallel}$ along and $s_{\perp}$ perpendicular to the
equipotential line. In these coordinates, an edge state that follows
the equipotential line at a small energy $E$ away from $E_F$ acquires
the additional phase $\Delta \phi = - l^{-2}_B \int ds \Delta
s_{\perp}$ with $e (dU/ds_{\perp}) \Delta s_{\perp}=E$. The potential
gradient $dU/ds_{\perp}$ determines the local electric field
$F(s)=-dU/ds_{\perp}$ at $s$. But $eF(s)l_B=\hbar v_{D}(s)$ where
$v_{D}(s)=F(s)/B$ is the drift velocity of the guiding center of
the cyclotron orbit at point $s$ of the edge state. Thus a small
increase in energy leads to a phase increment given by $\Delta \phi_i
= \int ds [1/\hbar v_D(s_i)]E$. A rough estimate using a drift
velocity which is constant along the edge gives $\Delta \phi_ i
\approx (L_i/\hbar v_D)E$ with $L_i$ the length of the edge state
$i$. For the phase-difference of the two interfering paths we have
\begin{equation}
\phi_1(E)-\phi_2(E)=\Delta \phi (E_F)+ E/E_c
\end{equation}
with $\Delta \phi (E_F )=\phi_{1}(E_F )-\phi_{2}(E_F )$ the
equilibrium phase difference. Formally, higher order terms in energy
can be neglected for characteristic energies $k_BT$ and $eV$ much smaller
than $(dU/ds_{\perp})^2/[d^2U/ds_{\perp}^2]$. The asymmetry of the two
edges thus gives rise to an energy scale $E_c=\{\int ds [1/\hbar v_D
(s_1 )] - \int ds [1/\hbar v_D(s_2)]\}^{-1}$ which is due to the
mismatch of the edge state path lengths, i.e. $E_c\approx \hbar v_D
/(\Delta L)$ with $\Delta L = L_1-L_2$. In principle, for a completely
symmetric interferometer one has $E_{c}\rightarrow \infty$.

\begin{figure}[tp]
\includegraphics[width=0.35 \textwidth,angle=270]{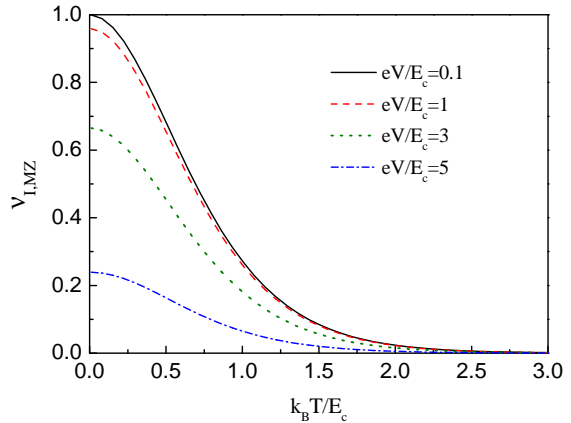}
\caption{Current visibility of the Mach-Zehnder interferometer 
$\protect\nu _{I,MZ}$ versus normalized
temperature $k_{B}T/E_{c}$ for $\mathcal{T}_{A}=\mathcal{T}_{B}.$}
\label{figcurrvisMZ}
\end{figure}

Given the scattering amplitudes $s_{\alpha \beta }$, the spectral
current density is found from Eqs. (\ref{Adef}), (\ref{currspecdens})
and (\ref{speccond}). For e.g. terminal 4, one gets
\begin{eqnarray}
j_{4}\left( E\right) &=&\left( e/h\right) \left[ f(E)-f_{0}(E)\right] \left[ 
\mathcal{T}_{A}R_{B}+\mathcal{T}_{B}R_{A}\right.  \notag \\
&&\left.+2\sqrt{\mathcal{T}_{A}\mathcal{T}_{B}R_{A}R_{B}}\cos \left(
E/E_{c}+\Theta \right) \right] ,
\end{eqnarray}
where we introduce the total, energy independent phase $\Theta
= \Delta \phi (E_F )+ 2\pi\Phi/\Phi_{0}$. Here $f_{0}(E)$ is the
distribution functions of the grounded terminals 2,3 and 4 and
$f(E)=f_{0}(E-eV)$ the distribution function of terminal 1. The
current is then given from Eq. (\ref{avcurr}), as
\begin{eqnarray}
&&I_{4}=\frac{e}{h}\left[ \left( \mathcal{T}_{A}R_{B}+\mathcal{T}%
_{B}R_{A}\right) eV+\sqrt{\mathcal{T}_{A}\mathcal{T}_{B}R_{A}R_{B}}\right. 
\notag \\
&\times&\left.4\pi k_{B}T\mathrm{csch}\left( \frac{k_{B}T\pi }{E_{c}}%
\right) \sin \left( \frac{eV}{2E_{c}}\right) \cos \left( \frac{eV}{2E_{c}}%
+\Theta \right) \right] .  \notag \\
&&
\label{currMZ}
\end{eqnarray}
Current conservation gives $I_{3}=(e^{2}/h)V-I_{4}$. The current
consists of two physically distinct parts. The first term in Eq. (\ref{currMZ})
is the phase independent, incoherent part, the current in the absence of
interference, while the second, phase dependent term is the
interference contribution. We note that a bias $eV$ of the order of
the asymmetry energy $E_{c}$ leads to the phase shifts of the
oscillation. The strength of the interference can conveniently be
quantified via the visibility as
\begin{equation}
\nu_{I}=\frac{I_{\max}-I_{\min }}{I_{\max}+I_{\min}}=\frac{\mathrm{amp}
[I]}{\left\langle I\right\rangle},
\end{equation}
which gives for the current in the MZ-interferometer
\begin{eqnarray}
\nu _{I,MZ} &=&\frac{\sqrt{\mathcal{T}_{A}\mathcal{T}_{B}R_{A}R_{B}}}{
\mathcal{T}_{A}R_{B}+\mathcal{T}_{B}R_{A}}  \notag \\
&\times& \frac{4\pi k_{B}T}{eV}\mathrm{csch}\left( \frac{k_{B}T\pi }{E_{c}}%
\right) \left| \sin \left( \frac{eV}{2E_{c}}\right) \right|.
\label{currvisMZ}
\end{eqnarray}
The visibility is a product of a term containing the QPC scattering
probabilities and a function depending on the energy scales $k_{B}T,$
$eV$ and $E_{c}$. The scattering probability term is maximum for
identical QPC's, $\mathcal{T}_{A}=\mathcal{T}_{B}.$ The energy scale
dependence is shown in Fig. \ref{figcurrvisMZ} where the visibility
for identical point contacts is plotted as a function of the
normalized temperature, $k_{B}T/E_{c}$. We note several interesting
features from Fig. \ref{figcurrvisMZ} and Eq. (\ref{currvisMZ}). (i)
the visibility shows decaying oscillations as a function of voltage
$\nu _{I,MZ}\varpropto \left| \sin \left( eV/2E_{c}\right) \right|
/eV$\ for arbitrary temperature. (ii) A symmetric MZ-interferometer,
$E_{c}\gg k_{B}T,$ $eV$, has unity visibility (for
$\mathcal{T}_{A}=\mathcal{T}_{B}$), i.e. shows perfect
interference. (iii) The visibility decays monotonically with
increasing temperature. For large temperatures, $k_{B}T\gg E_{c}$, the
visibility decays exponentially with the temperature as $\nu
_{I,MZ}\varpropto k_{B}T\exp \left( -\pi k_{B}T/E_{c}\right) .$

It is interesting to compare the calculated visibility to the
experimentally measured one in Ref. [\onlinecite{Ji03}]. As already
shown in Ref [\onlinecite{Ji03}], the measured scattering probability
dependence of $\nu_{I,MZ}$ is well reproduced by
Eq. (\ref{currvisMZ}). For the energy scale dependence, no information
about the drift velocity $v_D$ or the asymmetry $\Delta L$ needed to
determine $E_c$ is provided in Ref. [\onlinecite{Ji03}]. However, to
obtain the order of magnitude of $E_c$, considering as a rough
estimate a typical drift velocity \cite{driftrefs} $v_D\sim 10^4$ m/s
at a magnetic field $B \sim 1T$ and an asymmetry $\Delta L \sim 0.1
\mu $m gives an $E_c$ corresponding to an applied bias $\sim 10 \mu V$
or a temperature $\sim 100 mK$. These values are typically of the same
order of magnitude as the ones considered in the experiment. As a
first approximation, one would thus expect asymmetry effects to be of
importance. The observed temperature dependence, a strong decrease of
the visibility for increased temperature, is also qualitatively
described by Eq. (\ref{currvisMZ}) with an $E_c/k_B \sim 50$ mK. This
is however not the case with the voltage dependence. Ji et al find a
differential visibility, i.e. the visibility of $dI(V)/dV$, which
decays strongly with applied voltage, while Eq. (\ref{currMZ})
predicts a constant, voltage independent differential
visibility. There are several possible explanations to why the voltage
dependence in contrast to the temperature dependence is not reproduced
by the theory. Ji et al themselves point out two voltage dependent
dephasing mechanism: low frequency noise of $1/f$ type due to moving
impurities, induced by a higher current and fast fluctuations of the
potential landscape (and hence of the phase via the enclosed area)
caused by screening of the additional charges injected at higher
current. Screening might also, for the nonlinear current-voltage
characteristics predicted by Eq. (\ref{currMZ}), lead to a voltage
dependent renormalization of the transmission probabilities,
introducing a voltage dependence in the differential visibility.
\cite{tcmb,tcmb93} We also note that in the model of
Ref. [\onlinecite{Marquardt04c}], inducing dephasing by coupling the
MZ-interferometer to a quantum bath, gives a dephasing rate that
increases with increasing voltage. Clearly, further investigations are
needed to clarify the origin of the dephasing in the experiment in
Ref. [\onlinecite{Ji03}].

Turning to the noise, we focus on the cross correlator between
currents flowing in terminals 3 and 4 (the auto-correlator can be
obtained analogously). This allows for a straightforward comparison to
the result of the HBT-interferometer, for which the cross correlator
was investigated in Ref. [\onlinecite{Sam04}]. From Eqs. (\ref{noise})
and (\ref{noisedens}) and the expressions for the scattering
amplitudes, we arrive at the noise spectral density
\begin{eqnarray}
&&S_{34}\left( E\right) =\frac{-2e^{2}}{h}\left[ f(E)-f_{0}(E)\right]
^{2} \nonumber \\ &\times&\left\{ c_{0}+c_{\Theta }\cos \left(
\frac{E}{E_{c}}+\Theta \right)+c_{2\Theta }\cos \left(
2\left[\frac{E}{E_{c}}+\Theta \right] \right) \right\}, \nonumber \\
&&
\label{noisespecdens}
\end{eqnarray}
with coefficients 
\begin{eqnarray}
c_{0} &=&\mathcal{T}_{A}R_{A}+\mathcal{T}_{B}R_{B}-6\mathcal{T}_{A}R_{A}
\mathcal{T}_{B}R_{B},  \notag \\
c_{\Theta } &=&2\left( \mathcal{T}_{A}-R_{A}\right) \left( \mathcal{T}
_{B}-R_{B}\right) \sqrt{\mathcal{T}_{A}\mathcal{T}_{B}R_{A}R_{B}},  \notag \\
c_{2\Theta } &=&2\mathcal{T}_{A}\mathcal{T}_{B}R_{A}R_{B}.
\label{ccoeff}
\end{eqnarray}
Performing the energy integrals in Eq. (\ref{noise}) we find for the cross correlator
\begin{eqnarray}
S_{34} &=&-\frac{2e^{2}}{h}\left\{ c_{0}\bar{S}_{0}+c_{\Theta }\bar{S}%
_{\Theta }\cos \left( \frac{eV}{2E_{c}}+\Theta \right) \right.  \notag \\
&&\left. +c_{2\Theta }\bar{S}_{2\Theta }\cos \left[ 2\left( \frac{eV}{2E_{c}}%
+\Theta \right) \right] \right\}
\label{MZnoisetot}
\end{eqnarray}
where we introduce the functions
\begin{equation}
\bar{S}_{0}=eV\coth \frac{eV}{2k_{B}T}-2k_{B}T,
\label{enfcn1}
\end{equation}
\begin{eqnarray}
\bar{S}_{\Theta } &=&2\pi k_{B}T\mathrm{csch}\left( \frac{\pi k_{B}T}{E_{c}}%
\right) \left[ \coth \left( \frac{eV}{2k_{B}T}\right) \right.  \notag \\
&&\left. \times \sin \left( \frac{eV}{2E_{c}}\right) -\frac{k_{B}T}{E_{c}}%
\cos \left( \frac{eV}{2E_{c}}\right) \right].
\label{enfcn3}
\end{eqnarray}
and
\begin{eqnarray}
\bar{S}_{2\Theta } &=&2\pi k_{B}T\mathrm{csch}\left( \frac{2\pi k_{B}T}{E_{c}%
}\right) \left[ \coth \left( \frac{eV}{2k_{B}T}\right) \right.  \notag \\
&&\left. \times \sin \left( \frac{eV}{E_{c}}\right) -\frac{2k_{B}T}{E_{c}}%
\cos \left( \frac{eV}{E_{c}}\right) \right] .
\label{enfcn2}
\end{eqnarray}
containing the dependence on the energy scales $eV,$ $k_{B}T$ and $E_{c}$.

Just as the current in Eq. (\ref{currMZ}), the noise consists of a
 phase independent, incoherent part and a phase dependent,
 interference part. However, in contrast to the current, the phase
 dependent part of the noise contains two terms with different periods
 in $\Theta$, corresponding to oscillations periodic in $h/e$ and
 $h/2e$.  These terms result from two-particle scattering processes
 which enclose the AB-flux one and two times respectively. Similarly
 to the current, the phase of the oscillations are shifted for a bias
 $eV$ of the order of the asymmetry energy $E_{c}$.

It is important to note that in the MZ (in contrast to the HBT)
interferometer, two particle and higher order scattering processes are
just products of single particle scattering processes. The full
distribution of current fluctuations\cite{Forster05} is thus a
function of single particle scattering probabilities only. In
particular, the noise spectral density $S_{34}(E)$ in
Eq. (\ref{noisespecdens}) is proportional to $-|s_{41}|^2|s_{31}|^2$,
i.e. partition noise\cite{Buttiker92} with phase dependent scattering
probabilities. As a consequence, the phase independent, incoherent
part of the noise can not be understood as partition noise from
incoherent single particle processes, i.e. $\langle |s_{41}|^2
\rangle_{inc}\langle |s_{31}|^2 \rangle_{inc}\neq \langle
|s_{41}|^2|s_{31}|^2 \rangle_{inc}$. This is formally clear since the
term proportional to $\cos^2\Theta=[1+\cos(2\Theta)]/2$, from two
coherent scattering processes, obviously contribute to the phase
independent part of the noise. As a consequence, as shown by Marquardt
and Bruder,\cite{Marquardt04a,Marquardt04b} a model \cite{Blanter00}
with a filled stream of classical particles injected from reservoir
$1$ correctly reproduces the incoherent part of the current but fails
to reproduce the incoherent part of the noise. In contrast, as found
in Ref. [\onlinecite{Marquardt04b}] and further discussed below, the
completely dephasing voltage probe model correctly reproduces the
incoherent part of both the current and the noise.

To quantify the strength of the oscillations we introduce two separate
quantities, $\nu _{N,MZ}^{\Theta}$\ and $\nu _{N,MZ}^{2\Theta}$,
here simply called visibilities, which in close analogy to the current
visibility in Eq. (\ref{currvisMZ}) are defined as the ratio of the amplitudes of
the noise oscillations and the average noise. They become
\begin{equation}
\nu_{N,MZ}^{\Theta }=\frac{\left| c_{\Theta }\bar{S}_{\Theta }\right| }{c_{0}\bar{S}_{0}}
\label{noisevisMZ1}
\end{equation}
and 
\begin{equation}
\nu_{N,MZ}^{2\Theta }=\frac{\left| c_{2\Theta }\bar{S}_{2\Theta }\right| }{c_{0}\bar{S}_{0}}.
\label{noisevisMZ2}
\end{equation}
\begin{figure}[tp]
\includegraphics[width=0.55 \textwidth,angle=270]{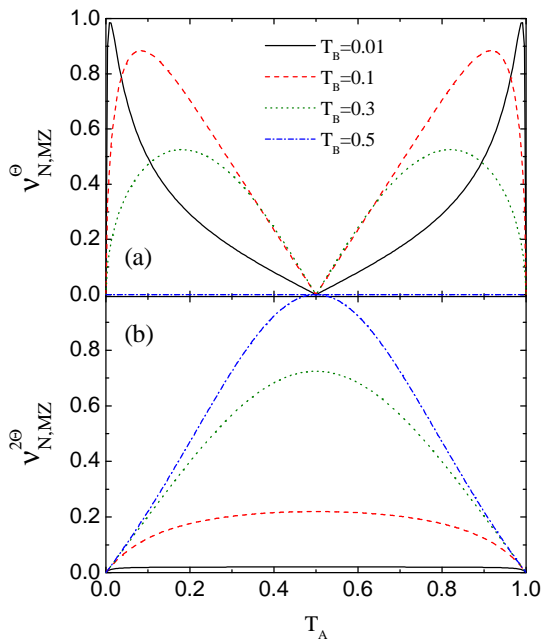}
\caption{Noise visibility $\nu_{N,MZ}^{\Theta}$ [figure
(a)] of the $h/e$ and $\nu_{N,MZ}^{2\Theta}$ [figure (b)] of the $h/2e$
oscillations in the shot noise of the Mach-Zehnder interferometer versus
transmission $\mathcal{T}_{A}$ of beam splitter $A$ for $E_{c}\gg
k_{B}T,eV$ for various transmission probabilities $\mathcal{T}_{B}$
of beam splitter $B$.}
\label{noisevis}
\end{figure}
Similarly to the current, both visibilities are products
of a term containing the scattering probabilities and a function of
the energy scales $eV$, $k_{B}T$ and $E_{c}$. We first focus on the
scattering probability dependent term by considering the visibility in
the limit of a symmetric interferometer, $E_{c}\gg eV$, $k_{B}T$,
where the energy-scale dependent terms are unity. This gives
\begin{equation}
\nu _{N,MZ}^{\Theta }=\frac{2\left| \left( \mathcal{T}_{A}-R_{A}\right)
\left( \mathcal{T}_{B}-R_{B}\right) \right| \sqrt{\mathcal{T}_{A}\mathcal{T}
_{B}R_{A}R_{B}}}{\mathcal{T}_{A}R_{A}+\mathcal{T}_{B}R_{B}-6\mathcal{T}
_{A}R_{A}\mathcal{T}_{B}R_{B}}
\end{equation}
and
\begin{equation}
\nu _{N,MZ}^{2\Theta
}=\frac{2\mathcal{T}_{A}\mathcal{T}_{B}R_{A}R_{B}}{\mathcal{T}_{A}R_{A}+\mathcal{T}_{B}R_{B}-6\mathcal{T}_{A}R_{A}\mathcal{T}
_{B}R_{B}}.
\end{equation}
The two visibilities are plotted in Fig. 4. Both visibilities are
symmetric under the substitutions $\mathcal{T}_{A}\leftrightarrow
R_{A}$ and $\mathcal{T}_{B}\leftrightarrow R_{B}$. The visibility $\nu
_{N,MZ}^{\Theta }$ is zero for $\mathcal{T}_{A}=R_{A}=1/2$, i.e. for
a symmetric setting of any of the QPC's. The visibility increases for
increasing QPC asymmetry, reaching a maximum for
$0<\mathcal{T}_{A}<0.5$\ and $0<\mathcal{T}_{B}<0.5$ (unity only in
the limit $\mathcal{T}_{A},\mathcal{T}_{B}\ll 1$) and then decreases
again toward zero at $\mathcal{T}_{A}=0$ or
$\mathcal{T}_{B}=0$. Interestingly, the visibility $\nu
_{N,MZ}^{2\Theta}$ shows an opposite behavior. It is maximum, equal to
unity, at $\mathcal{T}_{A}=\mathcal{T}_{B}=1/2$ and then decreases
monotonically for increasing QPC asymmetry, reaching zero at
$\mathcal{T}_{A}=0$ or $\mathcal{T}_{B}=0$. This different dependence
on the scattering probabilities makes it possible to investigate the
two oscillations independently by modulating the QPC transparencies.

Turning to the energy scale behavior, we consider for simplicity $\nu
_{N,MZ}^{\Theta }$ in the limit $\mathcal{T}_{A},\mathcal{T}_{B}\ll
1$\ and $\nu_{N,MZ}^{2\Theta }$ in the limit
$\mathcal{T}_{A}=\mathcal{T}_{B}=1/2$ where respective scattering
probability terms are unity. For a symmetric interferometer,
i. e. $E_{c}\gg eV,k_{B}T$, both visibilities are unity.  Considering
the situation when the temperature is comparable to the asymmetry
energy scale $E_{c}$ but the voltage is small $eV\ll k_{B}T,E_c$, we
get the visibilities
\begin{equation}
\nu _{N,MZ}^{\Theta }=\frac{\pi k_{B}T}{E_{c}}\mathrm{csch}\left( \frac{\pi
k_{B}T}{E_{c}}\right)\left[1+\left(\frac{k_BT}{Ec}\right)^2\right] 
\end{equation}
and
\begin{equation}
\nu _{N,MZ}^{2\Theta }=\frac{2\pi k_{B}T}{E_{c}}\mathrm{csch}\left( \frac{2\pi
k_{B}T}{E_{c}}\right)\left[1+4\left(\frac{k_BT}{Ec}\right)^2\right].
\end{equation}
\begin{figure}[t]
\includegraphics[width=0.45 \textwidth,angle=270]{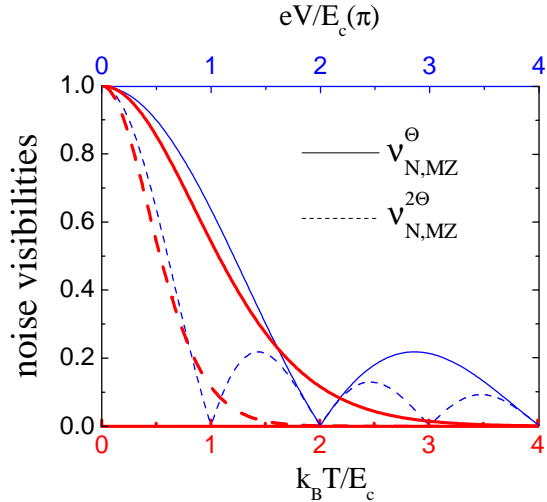}
\caption{Noise visibilities $\nu_{N,MZ}^{\Theta}$ (for
$\mathcal{T}_{A},\mathcal{T}_{B}\ll 1$) of the $h/e$ and $\nu_{N,MZ}^{2\Theta}$ of the 
$h/2e$ oscillations in the shot noise correlation of a Mach-Zehnder interferometer for
$\mathcal{T}_{A}=\mathcal{T}_{B}=1/2$ versus $k_{B}T/E_{c}$ for $eV\ll
k_{B}T,E_c$ (red curve) and versus $eV/E_{c} $ for $k_{B}T\ll E_{c},$
$eV$ (blue curve).}
\label{noisevis2}
\end{figure}
The temperature dependence of the visibilities are shown in
Fig. \ref{noisevis2}. Both visibilities decrease monotonically with
increasing temperature. For large temperature $k_{B}T\gg E_{c}$, the
visibilities decay exponentially as $\nu _{N,MZ}^{\Theta }\propto
(k_BT)^3\mbox{exp}\left( -\pi k_{B}T/E_{c}\right) $ and $\nu _{N,MZ}^{2\Theta
}\propto (k_BT)^3\mbox{exp}\left(-2\pi k_{B}T/E_{c}\right)$. \ The visibility
$\nu _{N,MZ}^{2\Theta }$ is thus considerably more sensitivity to
thermal smearing than $\nu _{N,MZ}^{\Theta }$. In the opposite limit,
for a small temperature but a voltage comparable to
$E_{c}$, i.e. $k_{B}T\ll E_{c},eV$, we instead get the visibilities
\begin{equation}
\nu _{N,MZ}^{\Theta}=\frac{2E_{c}}{eV}\left| \sin \left( \frac{eV}{2E_{c}}%
\right) \right|
\end{equation}
and 
\begin{equation}
\nu _{N,MZ}^{2\Theta}=\frac{E_{c}}{eV}\left| \sin \left( \frac{eV}{E_{c}}%
\right) \right| 
\end{equation}
The visibilities as a function of voltage are plotted in
Fig. \ref{noisevis2}. Both visibilities show an oscillating behavior,
decaying as a power law $\propto 1/eV$ with increasing voltage. The
period of oscillations, in $eV$, is $2\pi E_{c}$\ for $\nu
_{N,MZ}^{\Theta}$ but $\pi E_{c}$\ for $\nu _{N,MZ}^{2\Theta }$, half
the value for $\nu _{N,MZ}^{\Theta }$. The different voltage
dependence gives an additional possibility to investigate the two
visibilities independently.

In the experiment of Ji et al.\cite{Ji03} the noise was measured in
the high voltage regime, with the interference terms in both the
current and noise completely suppressed. The dependence of
the incoherent noise on the transparencies $\mathcal{T_{A}}$ and
$\mathcal{T_{B}}$ was investigated ($\mathcal{T_{A}}$ was kept at
$1/2$ and $\mathcal{T_{B}}$ was varied). A good agreement was found
with the first, incoherent term in Eq. (\ref{MZnoisetot}). Taken the
open questions on the effect of decoherence on the average current, a
detailed experimental investigation on the phase dependent,
interference part of the noise would be of great interest.

\subsection{Effect of dephasing}

\begin{figure}[t]
\includegraphics[width=0.45 \textwidth,angle=0]{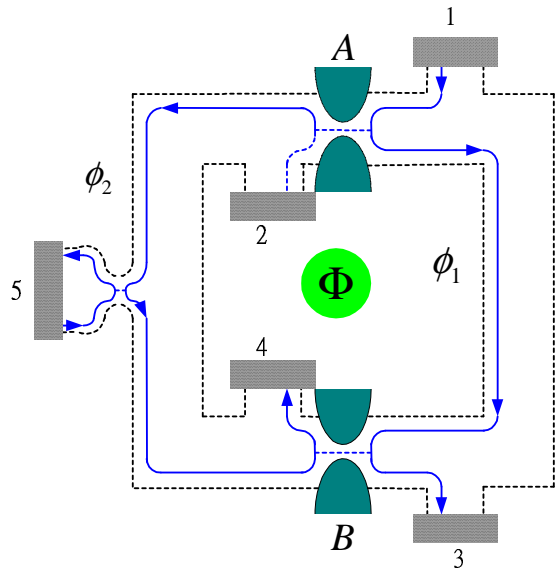}
\caption{The electrical MZ-interferometer, Fig. \ref{MZel}, with a dephasing voltage probe,
5, attached along one edge.}
\label{MZdeph}
\end{figure}

Next we consider the effect of dephasing on the current and noise.  As
discussed above, dephasing is introduced by connecting one of the two
arms of the interferometer to a fictitious, dephasing voltage
probe. The interferometer with the probe, denoted terminal $5$, is
shown in Fig. \ref{MZdeph}. The dephasing probe is connected to the
edge via a contact described by a scattering matrix
\begin{equation}
\left( \begin{array}{cc}
\sqrt{1-\varepsilon } & i\sqrt{\varepsilon } \\ 
i\sqrt{\varepsilon } & \sqrt{1-\varepsilon }
\end{array}\right),
\end{equation}
where the dephasing parameter $\varepsilon$ varies between $0$ (no
dephasing, fully coherent transport) and $1$ (complete dephasing,
fully incoherent transport). The presence of the dephasing probe
modifies the amplitudes for scattering between the terminal 1, 2, 3\
and 4. As an example, the scattering amplitude $s_{41}$, given in
Eq. (\ref{s41}) in the absence of dephasing, now becomes
\begin{eqnarray}
s_{41}\left( \varepsilon \right) &=&i\sqrt{\mathcal{T}_{B}R_{A}}e^{i(\phi
_{1}+\psi _{1})} \nonumber \\
&+&i\sqrt{1-\varepsilon}\sqrt{\mathcal{T}
_{A}R_{B}}e^{i(\phi_{2}-\psi _{2})}.
\end{eqnarray}

In addition, amplitudes for scattering into and out from the probe
terminal 5 have to be considered. The current is obtained from
Eqs. (\ref{avcurr}), (\ref{currspecdens}) and (\ref{fav}). For the
current in terminal 4, we find
\begin{eqnarray}
I_{4}^{dp}&=&\frac{e}{h}\left[\left(\mathcal{T}_{A}R_{B}+\mathcal{T}_{B}R_{A}\right)
eV+\right. \notag \\ 
&&\times
\sqrt{1-\varepsilon}\sqrt{\mathcal{T}_{A}\mathcal{T}_{B}R_{A}R_{B}} 4\pi
k_{B}T\mathrm{csch}\left(\frac{k_{B}T\pi}{E_{c}}\right)\notag \\
&&\times \left. \sin \left( \frac{eV}{2E_{c}}\right) \cos
\left(\frac{eV}{2E_{c}}+\Theta \right) \right].
\end{eqnarray}
Comparison with the result in the absence of dephasing in
Eq. (\ref{currMZ}) shows that the effect of the dephasing is to
suppress the phase-dependent oscillations by multiplying the
phase-dependent interference term with a factor
$\sqrt{1-\varepsilon}$. For complete dephasing $\varepsilon=1$, the phase
dependent term is completely suppressed. The effect of dephasing can
thus be simply incorporated in the visibility as
\begin{equation}
\nu _{I,MZ}^{dp}=\sqrt{1-\varepsilon}~\nu_{I,MZ},
\end{equation}
where $\nu _{I,MZ}$ is the visibility of the current oscillations in
the absence of dephasing, given by Eq. (\ref{currvisMZ}). As is clear
from the discussion above, to account for the experimental
observations in Ref. [\onlinecite{Ji03}], one would have to consider a
voltage dependent dephasing parameter $\varepsilon$.

Turning to the noise, we obtain the cross correlator between currents
in lead 3 and 4 in the presence of dephasing from Eqs. (\ref{noise})
and (\ref{dephnoisedens}), giving
\begin{eqnarray}
S_{34}^{dp}&=&-\frac{2e^{2}}{h}\left\{c_{0}\bar{S}_{0}+c_{\Theta
}\bar{S}_{\Theta }\sqrt{1-\varepsilon}\cos
\left(\frac{eV}{2E_{c}}+\Theta \right)\right. \notag \\
&&\left. +c_{2\Theta }\bar{S}_{2\Theta}\left(1-\varepsilon \right)
\cos\left[ 2\left( \frac{eV}{2E_{c}}+\Theta \right) \right] \right\}.
\end{eqnarray}
Here the terms $c_{0},$ $c_{\Theta},$
$c_{2\Theta},\bar{S}_{0},\bar{S}_{\Theta}$ and $\bar{S}_{2\Theta}$ are
defined above in Eqs. (\ref{ccoeff}) and (\ref{enfcn1}) to
(\ref{enfcn2}). Similarly to the current, the effect of the dephasing
is only to suppress the amplitude of the phase-dependent
oscillations. That is what one would naively expect to be the
consequence of pure dephasing. The two phase-dependent terms are
however affected differently, the $\cos \Theta$ term is suppressed by
a factor $\sqrt{1-\varepsilon}$ while the $\cos 2\Theta $ term is
suppressed by $\left( 1-\varepsilon \right) $.  The $\cos 2\Theta $
oscillations are thus more strongly suppressed. The visibilities of
the two oscillations in the presence of dephasing can simply be
written
\begin{equation}
\nu _{N,MZ}^{\Theta ,dp}=\sqrt{1-\varepsilon}\nu
_{N,MZ}^{\Theta }
\label{dephvisMZ1}
\end{equation}
and
\begin{equation}
\nu _{N,MZ}^{2\Theta ,dp}=(1-\varepsilon )\nu _{N,MZ}^{2\Theta },
\label{dephvisMZ2}
\end{equation}
where $\nu _{N,MZ}^{\Theta }$\ and $\nu_{N,MZ}^{2\Theta }$ are the
visibilities for the noise oscillations in the absence of dephasing,
given by Eqs. (\ref{noisevisMZ1}) and (\ref{noisevisMZ2}),
respectively.

Importantly, both oscillating terms are fully suppressed for complete
dephasing, $\varepsilon=1$.  Complete dephasing within the voltage
probe model thus gives a noise expression that only consists of the
phase independent, incoherent term in Eq.  (\ref{noisespecdens}). We
note already here that the same result is found below for the
HBT-interferometer. Since quantum interference by definition is
excluded from the model, i.e. all scattering phases are neglected, the
completely dephasing voltage probe thus constitutes a classical model
that correctly reproduces the incoherent part of the noise. As pointed
out above, a more detailed discussion of the physics of the voltage
probe and a comparison with
Refs. [\onlinecite{Marquardt04a,Marquardt04b,Clerk04}] is deferred to
a later work.

It is interesting to note that the effect of dephasing introduced with
the voltage probe, both for the current and noise, is for arbitrary
dephasing strength identical to a phase averaging. The result in
Eqs. (\ref{dephvisMZ1}) and (\ref{dephvisMZ2}) can be obtained by
averaging the fully coherent expressions in Eqs. (\ref{noisevisMZ1})
and (\ref{noisevisMZ2}) with respect to a Lorentzian distribution
$\rho(\Theta)$ of slow fluctuations of the phase $\Theta$ around the
average value $\Theta_0$, as
\begin{equation}
\int d\Theta \rho (\Theta) \cos \left( n\Theta
\right)=(1-\varepsilon)^{n/2}\cos \left(n\Theta_{0}\right).
\label{dephrel}
\end{equation}
with the Lorentzian distribution
\begin{equation}
\rho(\Theta)=\frac{a/\pi}{(\Theta-\Theta_0)^2+a^2}, \hspace{0.3cm} a=-(1/2)\mbox{ln}(1-\varepsilon)
\end{equation}
We note that, as pointed out in Ref. [\onlinecite{Marquardt04b}], a
Gaussian distribution of the phase fluctuations gives a different
result, not consistent with the dephasing voltage probe approach for
arbitrary dephasing stregth.

We emphasize that the results above are independent on to which edge
the probe is connected. Moreover, we also point out that the effect of
the voltage probes, for arbitrary $\varepsilon$, is multiplicative,
i.e. attaching $n$ voltage probes at arbitrary places along the arms
can be described by renormalizing $1-\varepsilon\rightarrow
(1-\varepsilon)^{n}$. Writing
$(1-\varepsilon)^{n}=\mbox{exp}(n\mbox{ln}[1-\varepsilon])=\mbox{exp}(-L/L_{\phi})$,
with $L_{\phi}=-d/\mbox{ln}[1-\epsilon]$ and $L=nd$ with $d$ the
distance between two probes, we can quite naturally incorporate the
effect of a uniform distribution of probes into a dephasing length
$L_{\phi}$. The suppression of the visibilities of the $h/e$ and
$h/2e$ oscialltions due to dephasing in Eqs. (\ref{dephvisMZ1}) and
(\ref{dephvisMZ2}) are then modified as
$(1-\varepsilon)^{1/2}\rightarrow \mbox{exp}(-L/2L_{\phi})$ and
$(1-\varepsilon)\rightarrow \mbox{exp}(-L/L_{\phi})$

\section{ Hanbury Brown Twiss interferometers}

The HBT-interferometer is less well known than the MZ-interferometer
and deserves some additional comments.\cite{HBTbook} The
HBT-interferometer was invented as a tool to measure the angular
diameter of stars. The first measurement \cite{HBT1} was carried out
on a radio star in 1954. Compared to existing schemes based on
Michelson interferometers, the HBT-interferometer proved to be less
sensitive to atmospheric scintillations, which allowed for a more
accurate determination of the angular diameter. After having
demonstrated a table-top version of the interferometer in the visual
range,\cite{HBT2} the angular diameter of the visual star Sirius was
determined.\cite{HBT3}

The experimental results, both the two-particle interference and the
positive intensity cross correlations, were successfully explained
within a semi-classical framework. Soon after the experiments, it was
however shown by Purcell\cite{Purcell56} that the positive cross
correlations could be explained in terms of bunching of individual
photons, emerging from the star, a thermal source of light. This
bunching was also demonstrated explicitly in subsequent photo counting
experiments.\cite{Twiss,Rebka} The HBT experiment thus laid the
foundations for quantum statistical methods in quantum
optics.\cite{Loudon} The HBT approach has also been of importance in
experimental particle physics.\cite{Baym} It is interesting to note
that positive intensity cross correlations between beams of light
emerging from a thermal source, according to some contemporary
\cite{Brannen,HBT4} ``{\it would call for a major revision of some
fundamental concepts in quantum mechanics}''.
Purcell,\cite{Purcell56} however, providing an elegant explanation of
the bunching phenomena, pointed out that ``{\it the Hanbury Brown
Twiss effect, far from requiring a revision of quantum mechanics, is
an instructive illustration of its elementary principles}''.

\begin{figure}[t]
\includegraphics[width=0.45 \textwidth,angle=0]{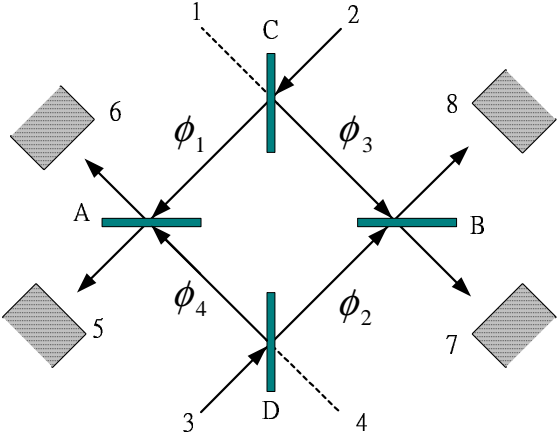}
\caption{Two-source, four-detector optical Hanbury Brown Twiss
geometry proposed in Ref. [\onlinecite{Sam04}].  Two beams of light incident from $2$ and $3$ are split in
partial beams at the semitransparent beam splitters $C$ and $D$. The
partial beams acquire geometrical phases $\phi_1$ to $\phi_4$ and are
rejoined in the beam splitters $A$ and $B$. The light intensity is
measured in detectors $5$ to $8$}
\label{HBTopt}
\end{figure}

An optical table-top version\cite{Yurke1,Yurke2} of the
HBT-interferometer is shown in Fig. \ref{HBTopt}. A beam of light is
emitted from each one of the sources $2$ and $3$, completely
uncorrelated with each other. The beams are split in two partial beams
at the semitransparent beam splitters $C$ and $D$ respectively. The
partial beams acquire phases $\phi_1$ to $\phi_4$ before scattering at
the second pair of beam splitters $A$ and $B$. The resulting beams are
collected in detectors at ports $5$ to $8$.

Importantly, there is no interference pattern in the intensities at
the detectors $5$ to $8$, instead the interference occurs only in the
cross correlations between intensities at $5,6$ and $7,8$. The
intensity cross correlations are sensitive to the two-particle
amplitudes: the interference is thus between two different
two-particle scattering events, e.g. (i) one particle from $2$
scatters to $5$ and one particle from $3$ scatters to $8$, with an
amplitude $A_1\mbox{exp}(i[\phi_1+\phi_2])$ and (ii) one particle from
$2$ scatters to $8$ and one particle from $3$ scatters to $5$, with an
amplitude $A_2\mbox{exp}(i[\phi_3+\phi_4])$. The amplitude to detect
one particle in $5$ and one in $8$ is then the sum of the two
two-particle amplitudes. This is the case since both scattering
processes have the same initial and final states and can not be
distinguished. The (reducible) cross correlation between intensities
in $5$ and $8$ is directly related to the corresponding two-particle
probability
$|A_1\mbox{exp}(i[\phi_1+\phi_2])+A_2\mbox{exp}(i[\phi_3+\phi_4])|^2$=
$|A_1|^2+|A_2|^2+2\mbox{Re}\left\{A_1A_2^*\exp(i[\phi_1+\phi_2-\phi_3-\phi_4])\right\}$. The
interference term
$2\mbox{Re}\left\{A_1A_2^*\exp(i[\phi_1+\phi_2-\phi_3-\phi_4])\right\}$
contains the four geometrical phases $\phi_1$ to $\phi_4$. The
HBT-interferometer is thus, in contrast to the MZ-interferometer, a
two-particle interferometer.
  
\begin{figure}[t]
\includegraphics[width=0.45 \textwidth,angle=0]{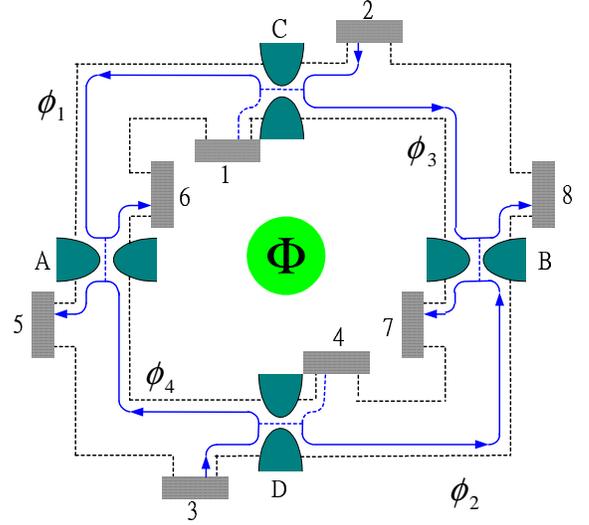}
\caption{Two-source, four-detector electrical Hanbury Brown Twiss
geometry implemented in a conductor in the Quantum Hall regime. The
electronic reservoirs $2$ and $3$ biased at $eV$ and reservoirs $1$
and $4$ to $8$ are kept at ground. The edge states (solid lines) have a
direction of transport indicated by arrows. The QPC's $A$ and $B$ play
the role of the beam splitters in Fig. \ref{HBTopt}. Geometrical phases
$\phi_1$ to $\phi_4$ and the AB-flux $\Phi$ are shown.}
\label{HBTel}
\end{figure}

The electrical analog of the HBT-interferometer, presented in
Ref. [\onlinecite{Sam04}], is shown in Fig. \ref{HBTel}. It consists of
a (rectangular) conductor with a hole in the middle, a Corbino
geometry. Similar to the MZ-interferometer, the electrons propagate
along single edge states. Scattering between the edge states take place
only at the beam splitters $A$ to $D$. The beam splitters are described
by scattering matrices given by Eq. (\ref{smat}). We first consider
the fully coherent case. In contrast to the MZ-interferometer, the
scattering amplitudes contain the phases $\phi _{i}$\ and $\psi _{i}$
only via multiplicative phase factors. As an example, the amplitude to
scatter from terminal $2$ to terminal 5 is given by
\begin{equation}
s_{52}=\sqrt{\mathcal{T}_{A}\mathcal{T}_{C}}e^{i(\phi_{1}-\psi _{1})}.
\label{s52}
\end{equation}
As a consequence, the average currents which depend only on the
modulus squared of the scattering amplitudes [see Eqs. (\ref{avcurr})
and (\ref{speccond})] do not contain any scattering phases. We get the
currents at terminals 5 to 8 as
\begin{eqnarray}
I_{5} &=&(e^{2}/h)V\text{ }\left( \mathcal{T}_{A}\mathcal{T}%
_{C}+R_{A}R_{D}\right), \notag \\ 
I_{6} &=&(e^{2}/h)V\text{
}\left(\mathcal{T}_{A}R_{D}+R_{A}\mathcal{T}_{C}\right),\text{ }
\notag \\ 
I_{7} &=&(e^{2}/h)V\text{ }\left(
\mathcal{T}_{B}R_{C}+R_{B}\mathcal{T} _{D}\right), \notag \\ 
I_{8} &=&(e^{2}/h)V\text{ }\left( \mathcal{T}_{B}\mathcal{T}
_{D}+R_{B}R_{C}\right).
\label{HBTcurr}
\end{eqnarray}
Turning to the current noise, the correlation between currents in
terminals 5,6 and 7,8 is given by Eq. (\ref{noisedens}). We find for
the spectral density for the correlators between terminal 5 and 8
\begin{eqnarray}
S_{58}\left( E\right) &=&\frac{-2e^{2}}{h}\left[ f(E)-f_{0}(E)\right] ^{2} 
\notag \\
&\times& \left\{c_{0,58}+c_{\Theta }\cos \left( E/E_{c}+\Theta \right)
\right\}
\end{eqnarray}
with the coefficients
\begin{eqnarray}
c_{0,58} &=&\mathcal{T}_{A}R_{B}\mathcal{T}_{C}R_{C}+\mathcal{T}_{B}R_{A}%
\mathcal{T}_{D}R_{D};  \notag \\
\bar c_{\Theta } &=&2\prod_{j=A,B,C,D}\sqrt{\mathcal{T}_{j}R_{j}},
\end{eqnarray}
and for the correlator between terminal 5 and 7  
\begin{eqnarray}
S_{57}\left( E\right) &=&\frac{-2e^{2}}{h}\left[ f(E)-f_{0}(E)\right] ^{2} 
\notag \\
&\times& \left\{ c_{0,57}+\bar c_{\Theta }\cos \left( E/E_{c}+\Theta \right)
\right\}
\end{eqnarray}
with the coefficient
\begin{equation}
c_{0,57}=\mathcal{T}_{A}\mathcal{T}_{B}\mathcal{T}_{C}R_{C}+R_{A}R_{B}%
\mathcal{T}_{D}R_{D}.
\end{equation}
Performing the energy integrals in Eq. (9), we obtain the corresponding
current cross correlators 
\begin{equation}
S_{58}=\frac{-2e^{2}}{h}\left[ c_{0,58}\bar{S}_{0}+\bar c_{\Theta }\bar{S}
_{\Theta }\cos \left( \frac{eV}{2E_{c}}+\Theta \right) \right]
\label{HBTnoise1}
\end{equation}
and
\begin{equation}
S_{57}=\frac{-2e^{2}}{h}\left[ c_{0,57}\bar{S}_{0}+\bar c_{\Theta }\bar{S}
_{\Theta }\cos \left( \frac{eV}{2E_{c}}+\Theta \right) \right] .
\label{HBTnoise2}
\end{equation}
Here $\bar{S}_{0}$ and $\bar{S}_{\Theta }$ are given by
Eqs. (\ref{enfcn1}) and (\ref{enfcn3}).  The other two correlators
$S_{67}$\ and $S_{68}$\ are given by the substitutions
$S_{67}=S_{58}\left( \mathcal{T}_{C}\leftrightarrow
\mathcal{T}_{D}\right) $\ and
$S_{68}=S_{57}\left(\mathcal{T}_{C}\leftrightarrow
\mathcal{T}_{D}\right) $.  Here, as for the MZ-interferometer we have
$\Theta = \Delta \phi (E_F ) + 2\pi \Phi /\Phi _{0}$ with $\Delta \phi
= \phi_{1}+ \phi_{2}-\phi_{3}-\phi_{4}$ and $\sum_{i=1}^{4}\psi
_{i}=2\pi \Phi /\Phi _{0}.$

Several observation can be made from the results above, put in
comparison with the result for the noise correlations for the
MZ-interferometer in Eq. (\ref{MZnoisetot}). Just as for the
MZ-interferometer, the noise consists of an incoherent, phase
independent part, and a coherent, interference part. The phase
dependent part of the noise in Eqs. (\ref{HBTnoise1}) and
(\ref{HBTnoise2}) however contains only one term. The amplitude of the
oscillating term is a product of a scattering probability term and an
energy-scale dependent function, just as for the
MZ-interferometer. This phase dependent term has the same dependence
on the phase $\Theta$, the same voltage dependent phase shift as well
as the same energy-scale dependence as the second term in
Eq. (\ref{MZnoisetot}). This is the case since they both arise from
processes which enclose the AB-flux once. Despite the fact that in the
HBT interferometer the AB-effect results from two-particle processes,
the periodicity is determined by the single electron flux quantum
$h/e$.  The dependence on the scattering probabilities is however
different, a consequence of the MZ and HBT interferometer geometries
being different. Importantly, there is no term in the noise in
Eqs. (\ref{HBTnoise1}) and (\ref{HBTnoise2}) that corresponds to the
last term in Eq. (\ref{MZnoisetot}), describing processes which
enclose the AB-flux twice. We note that the elementary scattering
processes in the HBT-interferometer, in contrast to the
MZ-interferometer, are two-particle processes. An important
consequence of this is that the incoherent, phase independent noise
term in Eqs. (\ref{HBTnoise1}) and (\ref{HBTnoise2}) can directly be
reproduced by a model with filled streams of classical particles
incident from reservoirs $2$ and $3$.
\begin{figure}[t]
\includegraphics[width=0.57 \textwidth,angle=270]{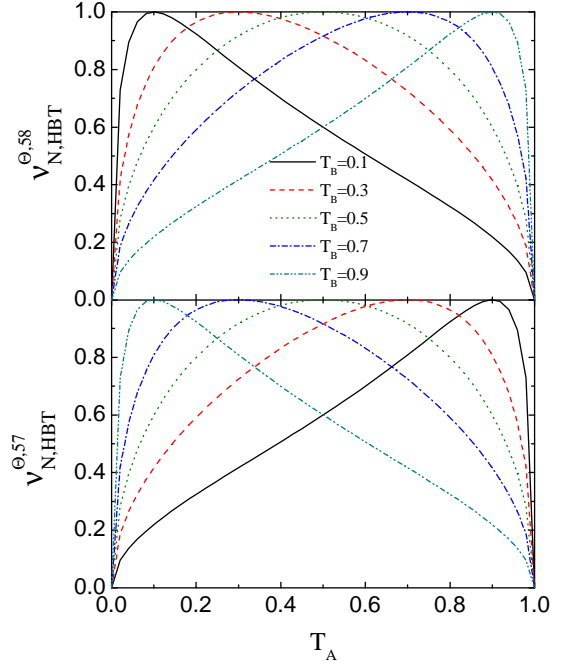}
\caption{Noise visibilities $\nu_{N,HBT}^{\Theta,58}$ and
$\nu_{N,HBT}^{\Theta,57}$ of shot noise correlations in the HBT
geometry versus transmission probability $\mathcal{T}_{A}$ for various
values of $\mathcal{T}_{B}$. A symmetric geometry, $E_c \gg kT,eV$,
and identical QPC's C and D are considered.}
\label{HBTnoisevisfig}
\end{figure}

Since there is only one phase-dependent term, the visibility of the
phase-dependent oscillations can again be directly defined, giving for
$\alpha=5,6$ and $\beta=7,8$
\begin{equation}
\nu_{N,HBT}^{\Theta,\alpha\beta}=\frac{\left|\bar{c}_{\Theta}\bar{S}_{\Theta}\right|
}{c_{0,\alpha\beta}\bar{S}_{0}}.
\end{equation}
Since the energy-scale dependence of the visibilities is identical to
$\nu_{N,MZ}^{\Theta }$ for the MZ-interferometer in
Eq. (\ref{noisevisMZ1}), shown in Fig. \ref{noisevis2}, we focus here
only on the scattering probability terms. We thus consider the limit
of a symmetric interferometer, $E_{c}\gg k_{B}T,eV$ for which the
energy-scale dependent part is unity. Several symmetries exists,
e.g. all visibilities $\nu_{N,HBT}^{\Theta,\alpha\beta}$ are unchanged
by the substitutions $R_C \leftrightarrow T_C$ and $R_D
\leftrightarrow T_D$. The visibility $\nu_{N,HBT}^{\Theta,58}$ is
unity for scattering probabilities obeying $T_AR_BR_CT_C=T_BR_AR_DT_D$
and similar relations hold for the other visibilities. All
visibilities go to zero for any of the transmission probabilities
approaching either zero or unity. Focusing on the case with $T_C=T_D$
(or equivalently $T_C=R_D$), the visibilities are given by
\begin{equation}
\nu _{N,HBT}^{\Theta ,58}=\nu _{N,HBT}^{\Theta ,67}=\frac{2\sqrt{\mathcal{T}
_{A}R_{A}\mathcal{T}_{B}R_{B}}}{\mathcal{T}_{A}R_{B}+\mathcal{T}_{B}R_{A}}
\end{equation}
and 
\begin{equation}
\nu _{N,HBT}^{\Theta ,57}=\nu _{N,HBT}^{\Theta ,68}=\frac{2\sqrt{\mathcal{T}%
_{A}R_{A}\mathcal{T}_{B}R_{B}}}{\mathcal{T}_{A}\mathcal{T}_{B}+R_{A}R_{B}}.
\end{equation}
The two different visibilities are plotted in
Fig. \ref{HBTnoisevisfig} as a function of $ \mathcal{T}_{A}$\ for
different $\mathcal{T}_{B}$. The visibility $\nu_{N,HBT}^{\Theta,58}$\
has a maximum equal to unity for $ \mathcal{T}_{A}=\mathcal{T}_{B}$,
while $\nu_{N,HBT}^{\Theta ,57}$\ instead has a maximum equal to unity
for $\mathcal{T}_{A}=R_{B}$.
\begin{figure}[t]
\includegraphics[width=0.45 \textwidth,angle=0]{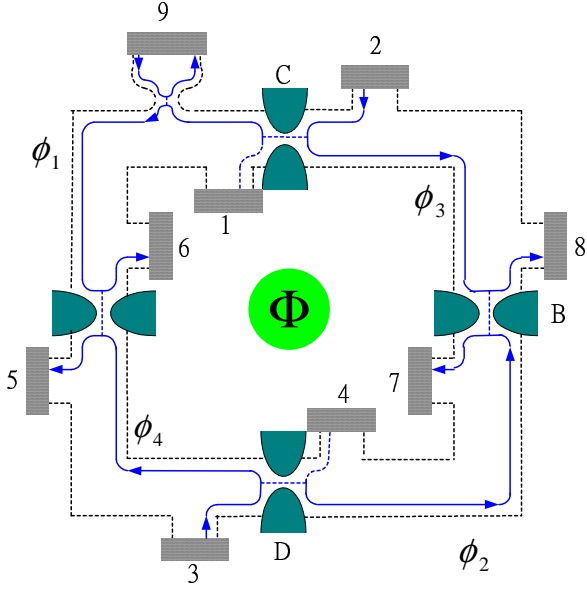}
\caption{The electrical HBT-interferometer, Fig. \ref{HBTel}, with a dephasing voltage probe,
$9$, attached along one edge.}
\label{HBTdepfig}
\end{figure}

\subsection{The effect of dephasing}

Just as in the MZ-interferometer, the dephasing in the
HBT-interferometer is introduced by connecting a fictitious voltage
probe to an edge between any of the two point contacts. The
HBT-interferometer with the probe, denoted 9, is shown in
Fig. \ref{HBTdepfig}. Here the probe is connected to the edge between
contact C and A, we emphasize that the results discussed below do not
depend on to which edge-state the probe is connected.

The presence of the probe modifies the amplitudes for scattering from
terminals 2, 3 to terminals 5 to 8. As an example, the scattering amplitude
in Eq. (\ref{s52}) is modified 
\begin{equation}
s_{52}=\sqrt{1-\varepsilon}\sqrt{\mathcal{T}_{A}\mathcal{T}_{C}}e^{i(\phi_{1}-\psi _{1})}.
\end{equation}
In addition, we also have to consider amplitudes for scattering into
and out from the probe terminal 9. The average currents in the
presence of dephasing, given from Eqs. (\ref{avcurr}) to
(\ref{speccond}) and (\ref{fav}), turn out to be given by the same
equations as in the absence of dephasing,
i.e. Eq. (\ref{HBTcurr}). This is what one expects, i.e. that
dephasing affects only the phase-dependent parts of the observables.

Turning to the current correlators, given from Eqs. (\ref{noise}),
(\ref{noisedens}) and (\ref{dephnoisedens}), we find for the
correlators between terminal 5 and 8
\begin{equation}
S_{58}^{dp}=\frac{-2e^{2}}{h}\left[ c_{0,58}\bar{S}_{0}+\bar{c}
_{\Theta }\bar{S}_{\Theta }\sqrt{1-\varepsilon}\cos \left(
\frac{eV}{2E_{c}}+\Theta \right) \right]
\label{HBTnoisedeph1}
\end{equation}
and for the correlators between terminals 5 and 7
\begin{equation}
S_{57}^{dp}=\frac{-2e^{2}}{h}\left[ c_{0,57}\bar{S}_{0}+\bar{c}
_{\Theta }\bar{S}_{\Theta }\sqrt{1-\varepsilon}\cos \left(
\frac{eV}{2E_{c}}+\Theta \right) \right].
\label{HBTnoisedeph2}
\end{equation}
The two remaining correlators are again given by the substitutions
$S_{67}=S_{58}\left(\mathcal{T}_{C}\leftrightarrow
\mathcal{T}_{D}\right) $\ and $S_{68}=S_{57}\left(
\mathcal{T}_{C}\leftrightarrow \mathcal{T}_{D}\right) $. We see from
Eq. (\ref{HBTnoisedeph1}) and (\ref{HBTnoisedeph2}) that just as for
the MZ-interferometer, the only effect of dephasing is to suppress the
phase-dependent term. The suppression factor is
$\sqrt{1-\varepsilon}$, just the same as for the $\cos \Theta $\ term
in the noise for the MZ-interferometer in Eq. (\ref{MZnoisetot}). We can
thus directly write the visibilities in the presence of dephasing as
\begin{equation}
\nu _{N,HBT}^{\Theta,\alpha \beta,dp}=\sqrt{1-\varepsilon}~\nu_{N,HBT}^{\Theta ,\alpha \beta }.
\end{equation}
This leads to the conclusion that the voltage probe for the
HBT-interferometer, just as for the MZ-interferometer, just has the
same effect as dephasing due to slow fluctuations of the phase $\Theta
$, with the distribution of the phase fluctuations obeying the
relation in Eq. (\ref{dephrel}). Moreover, the voltage probes have the
same multiplicative property as for the MZ-interferometer, allowing
one to describe the effect of a continuum of probes along the edges
(of total length $L=L_1+L_2+L_3+L_4$) with a dephasing length
$L_{\phi}$. The suppression of the visibilities of the $h/e$
oscillations due to dephasing are then modified as
$(1-\varepsilon)^{1/2}\rightarrow \mbox{exp}(-L/2L_{\phi})$, just as
for the $h/e$ oscillations of the MZ-interferometer.\\

\section{Conclusions}
The MZ-interferometer is an amplitude interferometer: it exhibits a
visibility in the average current with period $h/e$ and exhibits a
visibility in the shot noise with periods of both $h/e$ and $h/2e$. In
contrast, the HBT interferometer is an intensity interferometer, it
exhibits no AB-effect in the current and exhibits only an $h/e$-effect
in the shot noise correlations. Interestingly, our investigation shows
that the shot noise visibility of the HBT interferometer as a function
of temperature, voltage and dephasing rate, is qualitatively similar
to that of the $h/e$-component of the shot noise of the
MZ-interferometer. This is contrary to the naive expectation that the
visibility of the two particle processes which lead to the HBT effect
should be similar to the visibility of the two particle processes in
the MZ-interferometer, that is the $h/2e$ component of the shot noise.
Instead it is the number of times the AB-flux is enclosed which
determines the behavior of the visibility. 

In this paper we have investigated and compared in detail the voltage,
temperature and asymmetry dependence for the current and noise
visibilities in the MZ and HBT-interferometers. The experimental
realization of the HBT-interferometer is of large importance since it
allows for an unambiguous demonstration of two-particle interference
effects with electrons, to date not demonstrated. Moreover, a
successful realization of the HBT-interferometer would also enable a
first demonstration of orbital entanglement in electrical conductors,
a fundamentally important result. The results presented in this work
should prove useful for the experimental work aiming to detect the HBT
effect in electrical conductors.

\section{Acknowledgments}

We thank M. Heiblum, I. Neder, H. F\"orster and E. Sukhorukov for
stimulating discussions. This work was supported by the Graduate
Students Study Abroad Program, Taiwan National Science Council and the
Taiwan NSC93-2112-M-009-036, the Swedish Research Council and the
Swiss National Science Foundation and the network for Materials with
Novel Electronic Properties.

\end{document}